\documentstyle[aps,eqsecnum,tighten,epsf,preprint]{revtex}

\begin{document}

\draft

\preprint{\vbox {\hspace*{\fill} SNUTP-95/107 \\
          \hspace*{\fill} hep-ph/9510268}}

\title{Excited states of heavy baryons in the Skyrme model}

\author{Yongseok Oh\footnote{Address after January 1, 1996:
         Institute of Theoretical Physics,
         Physics Department, Technical University of Munich,
         D-85747 Garching, Germany.} }

\address{Department of Physics, National Taiwan University,
         Taipei, Taiwan 10764, Republic of China}

\author{Byung-Yoon Park\footnote{On leave of absence from
         Department of Physics, Chungnam National University,
         Daejeon 305-764, Korea.} }

\address{Institute of Nuclear Theory, University of Washington,
         Seattle, WA98195 \\
         and
         Department of Physics and Astronomy, University of
         South Carolina, Columbia, SC29208 }


\maketitle

\begin{abstract}
We obtain the spectra of excited heavy baryons containing one heavy
quark by quantizing the exactly-solved heavy meson bound states to
Skyrme soliton. The results are comparable to the recent experimental
observations and quark model predictions, and are consistent with the
heavy quark spin symmetry. However, somewhat large dependence of the
results on the heavy quark mass strongly calls for the incorporation
of the soliton-recoil effects.
\end{abstract}


\section{Introduction}

Up to the present, most ground state charm baryons containing one
$c$-quark, from $\Lambda_c^+$ to $\Omega_c^0$, have been observed
\cite{PDG94}. There have been much efforts to find excited charm
baryons and recently the experimental evidences for $\Sigma_c^*(2530)$
\cite{SKAT}, $\Lambda^*_c (2593)$ and $\Lambda^*_c (2625)$
\cite{ARGUS,E687,CLEO,CLEO94} are reported. Although their quantum
numbers are not identified yet, the spin-parity of the $\Sigma^*_c(2530)$
is interpreted as $j^\pi=\frac32{}^+$, and $\Lambda^*_c(2593)$ and
$\Lambda^*_c (2625)$ decaying to $\Lambda_c^+ \pi^+ \pi^-$ are regarded
as candidates for $j^\pi=\frac12{}^-$ and $\frac32{}^-$ excited states,
respectively, in accordance with the quark model predictions
\cite{CI,CIK,KT,KP,MI,Rosner}. The small mass splittings between
$\Sigma^*_c(2530)$ and $\Sigma_c(2453)$ and between $\Lambda_c^*(2593)$
and $\Lambda_c^*(2625)$ are consistent with the heavy quark spin symmetry
\cite{HQS,IW}, according to which the hadrons come in degenerate doublets
with total spin $j_\pm^{} = j_\ell^{} \pm \frac12$ (unless $j_\ell^{}$,
the total angular momentum of the light degrees of freedom, is zero)
in the limit of the heavy quark mass going to infinity.

On the other hand, the excited heavy baryons have been extensively studied
not only in various quark/bag models \cite{CI,CIK,KT,KP,MI,Rosner,IDS} but
also in heavy hadron chiral perturbation theory \cite{Cho} and in the bound
state approach of the Skyrme model \cite{RRS,OPM3,CW,SS94}. In the bound
state model, the heavy baryons are described by bound states of heavy
mesons and a soliton \cite{CK,BSHM}. A natural explanation of low-lying
$\Lambda(1405)$ is one of the success of the bound state approach
\cite{DNR}, where this $j^\pi = \frac12{}^-$ state is described by a
loosely bound $S$-wave $K$ meson to soliton. The same picture was
straightforwardly applied to the excited $\Lambda_c^* (\frac12{}^-)$ in
Ref. \cite{RRS}. The lack of the heavy quark symmetry in this first trial
is later supplied by treating the heavy vector mesons on the same footing
as the heavy pseudoscalar mesons \cite{BSHM}, and a generic structure of
the heavy baryon spectrum of orbitally excited states is established
\cite{OPM3}.

However, these works were done under the approximation that both the soliton
and the heavy mesons are infinitely heavy and so they sit on top of each
other. It is evident that this approximation cannot describe well the
orbitally and/or radially excited states due to the ignorance of any
kinetic effects. In Ref. \cite{CW}, the kinetic effects for the excited
states are estimated by approximating the static potentials for the heavy
mesons to the quadratic form with the curvature determined at the origin.
Such a harmonic oscillator approximation is valid only when the heavy
mesons are sufficiently massive so that their motions are restricted to
a very small range. The situation is improved in Ref. \cite{SS94} by making
an approximate Schr\"odinger-like equation and by incorporating the light
vector mesons. In a recent paper \cite{OP}, we have obtained all the energy
levels of the heavy meson bound states by solving {\it exactly\/}
the equations of motion from a given model Lagrangian without using any
approximations. (See also Refs. \cite{OPM1,SSVW}.) In this paper, we
quantize those states by following the standard collective coordinate
quantization method to investigate the excited heavy baryon spectra.

In the next section, we briefly describe our model Lagrangian and the
way of solving the equations of motion to obtain the bound states.
Then, in Sec. III, we quantize the soliton--heavy-meson bound system
based on the standard collective coordinate quantization method
and derive the mass formula. The resulting mass spectra of $\Lambda_c$,
$\Sigma_c$, $\Lambda_b$, and $\Sigma_b$ baryons are presented in Sec. IV
and compared with the recent experimental observations and with the
quark model predictions. Some detailed expressions are given in Appendix.

\section{The model}

We will work with a simple effective chiral Lagrangian of Goldstone
mesons and heavy mesons \cite{YCCLLY}:
\begin{equation}
\renewcommand\arraystretch{1.2} \begin{array}{rcl}
{\cal L} &=& {\cal L}^{\text{SM}}_{M}
          + D_\mu \Phi D^\mu \Phi^\dagger
          - m_\Phi^2 \Phi \Phi^\dagger
          - {\textstyle\frac12} \Phi^{*\mu\nu} \Phi^{*\dagger}_{\mu\nu}
          + m_{\Phi^*}^2 \Phi^{*\mu} \Phi^{*\dagger}_\mu
 \\
& & \mbox{} + f_Q ( \Phi A^\mu \Phi^{*\dagger}_\mu
                + \Phi^*_\mu A^\mu \Phi^\dagger )
          + {\textstyle\frac12} g_Q \varepsilon^{\mu\nu\lambda\rho}
            ( \Phi^*_{\mu\nu} A_\lambda \Phi^{*\dagger}_\rho
            + \Phi^*_\rho A_\lambda \Phi^{*\dagger}_{\mu\nu} ).
\end{array}\label{Lag} \end{equation}
Here, ${\cal L}^{\text{SM}}_{M}$ is an effective chiral Lagrangian of
Goldstone pions presented by an $SU(2)$ matrix field
$U=\exp(i\bbox{\tau}\cdot\bbox{\pi}/f_\pi)$,
which is simply taken as the Skyrme model Lagrangian,
\begin{equation}
{\cal L}^{\text{SM}}_{M} = \frac{f_\pi^2}{4} \,\text{Tr}\, (\partial_\mu
U^\dagger \partial^\mu U ) + \frac{1}{32e^2} \,\text{Tr}\, [ U^\dagger
\partial_\mu U, U^\dagger \partial_\nu U ]^2,
\end{equation}
with the pion decay constant $f_\pi$ and the Skyrme parameter $e$.
With the help of the Skyrme term, it supports a stable baryon-number-1
soliton solution under the hedgehog configuration
\begin{equation}
U_0 ( {\bf r} ) = \exp [ i \bbox{\tau} \cdot \hat {\bf r} F(r) ],
\end{equation}
where $F(r)$ satisfies the boundary conditions, $F(0)=\pi$ and
$F(\infty)=0$.

The heavy pseudoscalar and vector mesons containing a heavy quark $Q$
are represented by anti-isodoublet fields $\Phi$ and $\Phi^*_\mu$, with
their masses $m_\Phi^{}$ and $m^{}_{\Phi^*}$, respectively.
As an example, in case of $Q=c$, they are $D$ and $D^*$ meson
anti-doublets,
\begin{equation}
\Phi = (D^0, D^+) \quad \text{ and } \quad
\Phi^* = (D^{*0}, D^{*+}).
\end{equation}
The chiral transformations of the fields are defined as
\begin{equation}
\renewcommand\arraystretch{1.2} \begin{array}{l}
\xi \rightarrow L \xi h^\dagger = h \xi R^\dagger, \\
\Phi, \: \Phi^*_\mu \rightarrow \Phi h^\dagger, \: \Phi^*_\mu h^\dagger,
\end{array}\end{equation}
where $\xi \equiv \sqrt{U} = \exp(i\bbox{\tau}\cdot\bbox{\pi}/2f_\pi)$,
$L\in SU(2)_L$, $R\in SU(2)_R$, and $h$ is an $SU(2)$ matrix depending on
$L$, $R$, and $\xi$. The field $\xi$ defines vector and axial vector
fields as
\begin{equation} \renewcommand\arraystretch{1.5} \begin{array}{l}
V_\mu = {\textstyle\frac12} (\xi^\dagger \partial_\mu \xi
                           + \xi \partial_\mu \xi^\dagger), \\
A_\mu = {\textstyle\frac{i}{2}} (\xi^\dagger \partial_\mu \xi
                           - \xi \partial_\mu \xi^\dagger).
\end{array} \end{equation}
Then, the covariant derivatives are expressed in terms of $V_\mu$ as
\begin{equation}
D_\mu \Phi = \Phi ( \stackrel{\leftarrow}{\partial}_\mu + V_\mu^\dagger ),
\end{equation}
and a similar equation for $\Phi^*_\mu$, which defines the
field strength tensor of the heavy vector meson fields as
$ \Phi^*_{\mu\nu} = D_\mu \Phi_\nu^* - D_\nu \Phi_\mu^*$.

In our Lagrangian, we have a few parameters, $f_\pi$, $e$, $m_\Phi^{}$,
$m^{}_{\Phi^*}$, $f_Q$, and $g_Q$, which in principle have to be fixed
from the meson dynamics. We will use the experimental values for the
heavy meson masses. In order for the quantized soliton to fit the nucleon
and $\Delta$ masses \cite{ANW}, the pion decay constant has been adjusted
down to $f_\pi = 64.5$ MeV (with $e=5.45$). However, recently it is shown
that, taking into account the Casimir effect of the fluctuating pions
around the soliton configuration \cite{MK}, one can get reasonable nucleon
and $\Delta$ masses with the empirical value of the pion decay constant
($\sim$93 MeV). As for the coupling constants $f_Q$ and $g_Q$, there is no
sufficient experimental data to fix them. What is known to us is that, in
order for the Lagrangian to respect the heavy quark symmetry, they should
be related to each other as
\begin{equation}
\lim_{m_Q\rightarrow\infty}f_Q / 2 m_{\Phi^*}
= \lim_{m_Q\rightarrow\infty}g_Q = g,
\label{hqr}\end{equation}
and  the nonrelativistic quark model prediction on the universal constant
$g$ is $-0.75$ \cite{YCCLLY}, while the experimentally determined upper
bound is $g^2 \raisebox{-0.6ex}{$\stackrel{\textstyle{<}}{\sim}$} 0.5$
\cite{ACCMOR}. We will take $f_Q$ and $g_Q$ as free parameters with
keeping the relation (\ref{hqr}) and the nonrelativistic prediction
in mind.  The parameter dependence of the results will be discussed in
detail in Sec. IV. The Lagrangian (\ref{Lag}) is the simplest version
of the heavy meson effective Lagrangian and one may
include the light vector meson degrees of freedom such as $\rho$ and
$\omega$ \cite{SSVW,LVM} and the higher derivative terms to improve
the model predictions.

The equations of motion for the heavy mesons can be read off from the
Lagrangian (\ref{Lag}) as
\begin{equation} \renewcommand\arraystretch{1.5} \begin{array}{l}
D_\mu D^\mu \Phi^\dagger + m_\Phi^2 \Phi^\dagger = f_Q A^\mu
\Phi^{*\dagger}_\mu, \\
D_\mu \Phi^{*\mu\nu\dagger} + m_{\Phi^*}^2 \Phi^{*\nu\dagger} = - f_Q
A^\nu \Phi^\dagger + g_Q \varepsilon^{\mu\nu\lambda\rho} A_\lambda
\Phi^{*\dagger}_{\mu\rho},
\end{array} \label{eom}\end{equation}
with an auxiliary condition for the vector meson fields
\begin{equation}
m_{\Phi^*}^2 D_\nu \Phi^{*\nu\dagger} = - D_\nu D_\mu \Phi^{*\mu\nu\dagger}
- f_Q D_\nu ( A^\nu \Phi^\dagger ) + g_Q \varepsilon^{\mu\nu\lambda\rho}
D_\nu (A_\lambda \Phi^{*\dagger}_{\mu\rho}),
\label{ac}\end{equation}
which reduces to the Lorentz condition $\partial^\mu \Phi^{*\dagger}_\mu =0$
in case of the free vector meson fields. To avoid any unnecessary
complications associated with the anti-doublet structure of $\Phi$ and
$\Phi^*_\mu$, we work with $\Phi^\dagger$ and $\Phi^{*\dagger}_\mu$. When
the static hedgehog configuration of $U_0$ is substituted, the vector and
axial vector fields become
\begin{equation} \renewcommand\arraystretch{1.5} \begin{array}{l}
V^\mu = ( V^0, {\bf V} ) = \bbox{(} 0, -i ( \bbox{\tau} \times \hat
{\bf r} ) \upsilon (r) \bbox{)}, \\
A^\mu = ( A^0, {\bf A} ) = \bbox{(} 0, \frac12 [ a_1(r) \bbox{\tau}
      + a_2(r) \hat {\bf r} \bbox{\tau} \cdot \hat {\bf r} ] \bbox{)},
\end{array} \label{VA} \end{equation}
where
$\upsilon(r) = [ \sin^2 ({F}/{2}) ]/r $, $a_1 (r) = (\sin F) / r$ and
$a_2 (r) = F' - (\sin F)/ r$.

Then, the problem becomes to find the classical eigenmodes (especially
the bound states) of the heavy mesons moving under the static potentials
formed by the soliton field. The equations are invariant under parity
operations and the ``grand spin" rotation generated by the operator
\begin{equation}
{\bf K} = {\bf J} + {\bf I} = {\bf L} + {\bf S} + {\bf I},
\end{equation}
where ${\bf L}$, ${\bf S}$, and ${\bf I}$ are the orbital angular
momentum, spin, and isospin operator of the heavy mesons, respectively.
This allows us to classify the eigenstates by the grand spin quantum
numbers $(k,k_3)$ and the parity $\pi$. We will denote the set of quantum
numbers by $\{n\}$ $(\equiv \{\bar n;k,k_3;\pi\}$, $\bar n$ is a quantum
number to distinguish the radial excitations). For a given grand spin
$(k,k_3)$ with parity $\pi = (-1)^{k\pm 1/2}$, the general wave function
of an energy eigenmode can be written as
\begin{equation} \renewcommand\arraystretch{1.5}\begin{array}{l}
\displaystyle
\Phi^{\dagger}_n({\bf r},t) = e^{+i\varepsilon_n t} \varphi_n(r)
{\cal Y}^{(\pm)}_{kk_3}(\hat {\bf r}), \hskip 5mm
\Phi^{*\dagger}_{0,n}({\bf r},t) = e^{+i\varepsilon_n t} i
\varphi^*_{0,n}(r) {\cal Y}^{(\mp)}_{kk_3}(\hat {\bf r}), \\
\bbox{\Phi}^{*\dagger}_n({\bf r},t) = e^{+i\varepsilon_n t}
\left[ \varphi^*_{1,n}(r) \,
\hat{{\bf r}} {\cal Y}^{(\mp)}_{kk_3}(\hat {\bf r})
+ \varphi^*_{2,n}(r) \, {\bf L} {\cal Y}^{(\pm)}_{kk_3}(\hat {\bf r})
+ \varphi^*_{3,n}(r) \,
 {\bf G} {\cal Y}^{(\mp)}_{kk_3}(\hat {\bf r}) \right],
\end{array} \label{wf}\end{equation}
where ${\bf G} \equiv -i ( \hat {\bf r} \times {\bf L} )$ and ${\cal
Y}^{(\pm)}_{kk_3} (\hat {\bf r})$ is the (iso)spinor spherical harmonic
obtained by combining the eigenstates of ${\bf L}$ and ${\bf I}$.
The wave functions should be normalized so that the eigenmodes carry
a unit heavy flavor number ($C=+1$ and $B=-1$). The normalization condition
is given in Appendix. Note the different sign convention of the energy
in the exponent for the time evolution of the eigenmodes and that
$\varphi^*_3 (r)$ $[\varphi^*_2 (r)]$ is absent in case of
$k^\pi = \frac12{}^+(\frac12{}^-)$. Substituting Eq. (\ref{wf})
into the equations of motion (\ref{eom}) and auxiliary condition
(\ref{ac}), one can obtain coupled differential equations for the radial
functions, $\varphi (r)$ and $\varphi^*_\alpha (r)$ ($\alpha=0,\dots,3$).
(See Ref. \cite{OP} for more details.)

Given in Fig. 1 is a typical energy spectrum of bound heavy meson states
obtained by solving the equations numerically with experimental heavy
meson masses ($m_D = 1867$ MeV, $m_{D^*} = 2010$ MeV and $m_B = 5279$ MeV,
$m_{B^*} = 5325$ MeV), $f_\pi = 64.5$ MeV, $e=5.45$, and $f_Q/2m_{\Phi^*}
=g_Q = -0.75$. We present the binding energies defined as $\Delta
\varepsilon = m_{\Phi}^{} - \varepsilon$. Comparing it with the energy
spectrum obtained in the infinite heavy mass limit \cite{OPM3}, one can
see that the ``parity doubling" artifact is removed by the centrifugal
energy contribution and there appear many radially excited states. As a
trace of the heavy quark symmetry, the energy levels come in nearly
degenerate doublets with grand spin $k_\pm = k_\ell \pm \frac12$ (unless
$k_\ell = 0$) and parity $\pi=(-1)^{k_\ell}$ \cite{OPM3}, where ${\bf
K}_\ell \equiv {\bf K} - {\bf S}_Q$ with the heavy quark spin ${\bf
S}_Q$. The energy levels are obtained in the soliton-fixed frame, which
must be a crude approximation. The soliton-recoil effects should be
incorporated in order for the bound state approach to work well with
heavy flavors. In this work, however, we will proceed without incorporating
the soliton-recoil effects. We will discuss some possible modifications
of the results in Sec. IV, leaving the rigorous and detailed investigations
to our future study.

\section{Quantization}

The soliton--heavy-meson bound system described so far does not carry any
good quantum numbers except the grand spin, parity, and baryon number.
In order to describe baryons with definite spin and isospin quantum
numbers, we should quantize the system by going to the next order in
$1/N_c$ \cite{CK}. This can be done by introducing collective variables
to the zero modes associated with the invariance of the
soliton--heavy-meson bound system under simultaneous isospin rotation
of the soliton together with the heavy meson fields:
\begin{equation}\renewcommand\arraystretch{1.2}\begin{array}{c}
\xi({\bf r},t) = C(t) \, \xi_0({\bf r}) \, C^\dagger(t), \\
\Phi({\bf r},t) = \Phi_{\text{bf}}({\bf r},t) \, C^\dagger(t), \\
\Phi^*_\mu({\bf r},t) = \Phi^*_{\text{bf},\mu}({\bf r},t) \, C^\dagger(t),
\end{array}\label{cv}\end{equation}
where $\xi_0^2 \equiv U_0$ and $C(t)$ is an SU(2) matrix. The subscript
``bf" is to denote that they are the fields in the body-fixed (isospin
co-moving) frame. (Hereafter, we will drop it to shorten the notation and
all the heavy meson fields appearing in equations are those in the
body-fixed frame unless specified.) Assuming sufficiently slow collective
rotation, we will work in the Born-Oppenheimer approximation where the
bound heavy mesons remain in an unchanged classical eigenmode.

Introduction of the collective variables as Eq. (\ref{cv}) leads us to an
additional Lagrangian of $O(1/N_c)$,
\begin{equation}
L_{-1} = {\textstyle\frac12} {\cal I} \bbox{\omega}^2
       + \bbox{\omega} \cdot \bbox{\Theta},
\end{equation}
where the angular velocity $\bbox{\omega}$ of
the collective rotation is defined by
\begin{equation}
C^\dagger \dot C \equiv {\textstyle\frac{i}{2}} \bbox{\tau} \cdot
\bbox{\omega},
\end{equation}
and ${\cal I}$ is the moment of inertia of the soliton \cite{ANW}.
The explicit form of $\bbox{\Theta}$ is
\begin{equation}\renewcommand\arraystretch{1.2}\begin{array}{rcl}
\bbox{\Theta} &=&  \displaystyle\int d^3 r \Bigl\{
  {\textstyle \frac{i}{2}}
  ( \dot \Phi {\bf T}_V \Phi^\dagger - \Phi {\bf T}_V \dot \Phi^\dagger
    + \dot \Phi^{*i} {\bf T}_V \Phi^{*i\dagger}
    - \Phi^{*i} {\bf T}_V \dot \Phi^{*i\dagger} ) \\
&& \hskip 3em \mbox{} + {\textstyle\frac{i}{2}}
              [ \Phi^{*i} {\bf T}_V D^i \Phi^{*\dagger}_0
              - (D^i \Phi^*_0) {\bf T}_V \Phi^{*i\dagger} ]
 \\ &&\hskip 3em
\mbox{}  - {\textstyle\frac12} f_Q
              ( \Phi {\bf T}_A \Phi^{*\dagger}_0
              + \Phi^*_0 {\bf T}_A \Phi^\dagger )
  - {\textstyle\frac{i}{2}} g_Q \varepsilon^{ijk}
              \Phi^{*i} \{ {\bf T}_V, A^j \}_+ \Phi^{*k\dagger}
 \\ &&\hskip 3em
\mbox{}  - {\textstyle\frac12} g_Q \varepsilon^{ijk}
              [ (D^i \Phi^{*j}) {\bf T}_A \Phi^{*k\dagger}
              + \Phi^{*k} {\bf T}_A D^i \Phi^{*j\dagger} ]
\Bigr\},
\end{array}\label{pth}\end{equation}
where $\{A,B\}_+ \equiv AB + BA$ and
\begin{equation} \renewcommand\arraystretch{1.5} \begin{array}{l}
{\bf T}_V \equiv {\textstyle\frac12}
     ( \xi_0^\dagger \bbox{\tau} \xi_0
     + \xi_0 \bbox{\tau} \xi_0^\dagger )
= t_1(r) \bbox{\tau} + t_2(r) \hat {\bf r} \bbox{\tau} \cdot
  \hat {\bf r}, \\
{\bf T}_A \equiv {\textstyle\frac12}
     ( \xi_0^\dagger \bbox{\tau} \xi_0
     - \xi_0 \bbox{\tau} \xi_0^\dagger )
= t_3(r) (\bbox{\tau} \times \hat {\bf r}),
\end{array} \label{tva} \end{equation}
with $t_1(r) = \cos F$, $t_2(r) = 1 - \cos F$, and $t_3(r) = \sin F$.
Note that it is nothing but the isospin current of the heavy mesons interacting
with Goldstone bosons (modulo the sign) as discussed in Ref. \cite{OPM3}.

The spin operator ${\bf J}$ and isospin operator ${\bf I}$ of the
system can be obtained by applying the N\"{o}ther theorem to the
invariance of the Lagrangian under the corresponding rotations:
\begin{eqnarray} \renewcommand\arraystretch{1.5} \begin{array}{l}
I_a = D_{ab}(C) R_b, \\
{\bf J} = {\bf R} + {\bf K}_{\text{bf}},
\end{array}
\end{eqnarray}
where ${\bf R}$ is the rotor spin conjugate to the collective variables,
\begin{equation}
{\bf R} = \frac{\delta L}{\delta \bbox{\omega}}
        = {\cal I} \bbox{\omega} + \bbox{\Theta}.
\end{equation}
${\bf K}_{\text{bf}}$ is the grand spin operator of the heavy meson fields
(in the isospin co-moving frame) and $D_{ab}(C) [\equiv \frac12 \text{Tr}
(\tau_a C \tau_b C^\dagger)]$ is the adjoint representation of the
collective variables. Note that the grand spin operator plays the role of
the spin operator for the heavy mesons, that is, their isospin is
transmuted into a part of the spin.

The physical heavy baryon states with spin-parity $j^\pi$ and isospin $i$
can be obtained by combining the rotor spin eigenstates and the
heavy meson bound states of grand spin $k$ with the help of the
Clebsch-Gordan coefficients:
\begin{mathletters}  \label{state}
\begin{equation}
\mid i,i_3; j^\pi\!, j_3 \rangle\!\rangle = \sum_{k_3}
\langle i, j_3\!-\!k_3, k, k_3\! \mid j, j_3 \rangle
\mid i; i_3, j_3\!-\!k_3 \} \mid \bar n;k, k_3; \pi \rangle,
\end{equation}
with $k = |j-i|, |j-i|+1, \dots, i+j$. Here, $\mid i; m_1, m_2 \}$
($m_1,m_2 = -i, -i+1, \dots, i$) denotes the eigenstate of the
rotor-spin operator $R_a$:
\begin{equation} \begin{array}{l}
{\bf R}^2 | i; m_1, m_2 \} = i(i+1) \mid i; m_1,m_2\}, \\
R_z \mid i; m_1, m_2 \} = m_2 \mid i; m_1,m_2\}, \\
I_z \mid i; m_1, m_2 \} = m_1 \mid i; m_1,m_2\},
\end{array}
\end{equation}
and $\mid \bar n; k,k_3; \pi \rangle$ is the single-particle Fock state of the
heavy meson fields where one classical eigenmode of the corresponding
grand spin quantum number is occupied:
\begin{equation} \begin{array}{l}
{\bf K}^2_{\text{bf}} \mid \bar n; k, k_3; \pi \rangle = k(k+1) \mid \bar
n; k, k_3; \pi \rangle, \\
K_{\text{bf},z} \mid \bar n; k, k_3; \pi \rangle = k_3 \mid \bar n;
k, k_3; \pi \rangle.
\end{array} \end{equation}
As an artifact of large $N_c$ feature of the Skyrme model, one may have
baryon states with higher isospin $i\geq 2$. We will restrict our
considerations to the heavy baryons with $i=0$ $(\Lambda_Q^{})$ and
$i=1$ $(\Sigma_Q^{})$. To be precise, in Eq. (\ref{state}), one may have to
sum over the possible $k$ as
\end{mathletters}
\begin{equation}
\mid i,i_3; j^\pi\!, j_3 \rangle\!\rangle = \sum_{k,k_3} \alpha_k^{}
\langle i, j_3\!-\!k_3, k, k_3 \mid j, j_3 \rangle
\mid i; i_3, j_3\!-\!k_3 \} \mid \bar n; k, k_3; \pi \rangle,
\end{equation}
with the expansion coefficients $\alpha_k$ to be determined by
diagonalizing the Hamiltonian. However, as far as the heavy baryon
states are concerned, the mixing effects are rather small. It is shown
in Ref. \cite{OPM3} that there is no mixing even when two states become
degenerate in the $m_Q\rightarrow \infty$ limit. (Such a mixing effect
plays the most important role in establishing the heavy quark symmetry
in the pentaquark baryons \cite{OPM2}.) Thus, we will involve only one
single-particle Fock state in the combination (\ref{state}).

The physical baryons should be the eigenstates of the Hamiltonian and
their masses come out as eigenvalues. The Hamiltonian can be obtained
by taking the Legendre transformation with the collective variables and
the heavy meson fields taken as dynamical degrees of freedom. Up to the
order of $1/N_c$, we have
\begin{mathletters} \label{H}
\begin{equation}
H = H^{+1} + H^{0} + H^{-1},
\end{equation}
where $H^{m}(m=+1,0,-1)$ is the Hamiltonian of $O(N_c^m)$. The Hamiltonian
of the leading order in $N_c$ is the soliton mass \cite{ANW}:
\begin{equation}
H^{+1} = M_{\text{sol}},
\end{equation}
and $H^{0}$ is the Hamiltonian of the heavy meson fields which yields
the eigenenergy $\varepsilon_n^{}$ when acts on the single-particle
Fock state $\mid \{n \} \rangle$:
\begin{equation}
H^{0} \mid \{ n \} \rangle = \varepsilon_n \mid \{ n \} \rangle.
\end{equation}
Finally, the Hamiltonian of order of $1/N_c$ arising from the collective
rotation is in a form of
\begin{equation}
H^{-1} =  \frac{1}{2{\cal I}} ( {\bf R} - \bbox{\Theta} )^2.
\end{equation}

We will take the $1/N_c$ order term as a perturbation. Then, the mass of
the heavy baryon state (\ref{state}) is obtained as
\end{mathletters}
\begin{equation}
m_{(i,j^\pi)} = M_{\text{sol}} + \varepsilon_n^{} +
\frac{1}{2{\cal I}} \langle\!\langle i; j^\pi \mid (\bbox{R} - \bbox{\Theta})^2
\mid i; j^\pi \rangle\!\rangle,
\end{equation}
where $\varepsilon_n^{}$ is the eigenenergy of the heavy meson bound
state involved in the construction of the state $|i;j^\pi\rangle\!\rangle$.
If only one single-particle Fock state $\mid \bar n; k, k_3; \pi \rangle$
is involved in Eq. (\ref{state}), we can write the mass formula in a more
convenient form as
\begin{equation} \label{massform}
m_{(i,j^\pi)} = M_{\text{sol}} + \varepsilon_n^{} + \frac{3}{8{\cal I}}
         + \frac{1}{2{\cal I}} [ c_n^{} j(j+1)
          + (1-c_n^{}) i(i+1) - c_n^{} k(k+1) ].
\end{equation}
Here, we have used the Wigner-Eckart theorem to express
the expectation value of $\bbox{\Theta}$ as
\begin{equation}
\langle \bar n; k, k^\prime_3; \pi \mid \bbox{\Theta}
\mid \bar n;k, k_3; \pi \rangle \equiv -c_n^{}
\langle \bar n;k, k^\prime_3; \pi \mid {\bf K}_{\text{bf}}
\mid \bar n;k, k_3; \pi \rangle,
\label{theta}  \end{equation}
which defines the ``hyperfine splitting" constant $c_n$. The explicit
expressions for $c_n$ are given in Appendix. In evaluating the
expectation value of $\bbox{\Theta}^2$, we have used the fact that
$\bbox{\Theta}$ is the isospin operator (with opposite sign), which implies
\begin{equation}
\bbox{\Theta}^2 \mid n; k, k_3; \pi \rangle
 = \textstyle{\frac34} \mid n; k, k_3; \pi \rangle.
\end{equation}

\section{Results and Discussions}

As for the $\Lambda_Q$ baryons that are constructed with the $i=0$ rotor
spin state and one single-particle Fock state $\mid \bar n; k\!=\!j, \,
k_3; \pi \rangle$, the mass formula can be further simplified as
\begin{equation}
m_{\Lambda_Q(j)}^{} = M_{\text{sol}} + \varepsilon_n^{} + \frac{3}{8{\cal I}}
= m_N^{} + m_{\Phi} + \Delta\varepsilon_n^{},
\end{equation}
where $m_N^{}$ is the nucleon mass ($m_N^{} = M_{\text{sol}} + 3/8{\cal I}$)
and $\Delta\varepsilon_n^{} = \varepsilon_n - m_{\Phi}$. Thus, the mass
spectrum of $\Lambda_Q$ baryons is exactly the same as Fig. 1 with
replacing the $\Delta\varepsilon=0$ line by the $m_N + m_\Phi$ threshold.
However, the $\Lambda_Q$ spectrum obtained with the parameters of Fig. 1
(Set 1) is not at the level of being compared with experiments. As can be
seen in Fig. 1, the binding energy ($\sim 380$ MeV) and the mass splitting
($\sim 200$ MeV) between the first excited $\Lambda_c^*$ and the ground
state are too small compared with the experimental values, $\sim 520$ and
$\sim 310$ MeV, respectively. However, we can easily improve the situation
by adjusting the parameters within a reasonable range. Table I summarizes
the parameter sets that we will examine and the parameter dependence of
$\Lambda_c$ spectra are shown in Fig. 2.

What we want to have is more deeply bound states with wider level
splittings, which can be achieved if we have a deeper and narrower
interacting potentials in the equations of motion for the heavy mesons.
One way of obtaining such potentials in a given model Lagrangian is to
take the empirical value for $f_\pi$ instead of $f_\pi = 64.5$ MeV. Since
the soliton wave function $F(r)$ is only a function of a dimensionless
variable $x=ef_\pi r$ (in the chiral limit), the functions $a_1(r)$ and
$a_2(r)$ appearing in potentials scale with the factor $ef_\pi$.
Furthermore, the soliton mass $M_{\text{sol}}$ and the moment of
inertia ${\cal I}$ come out in the form of \cite{ANW}
\begin{equation}
M_{\text{sol}} = \tilde{M} f_\pi/e \quad\text{ and }\quad
{\cal I} = \tilde{{\cal I}}/(e^3f_\pi),
\end{equation}
with dimensionless quantities $\tilde{M}$ and $\tilde{{\cal I}}$ that are
independent of $e$ and $f_\pi$. If we are to have a correct $\Delta$-$N$
mass splitting ($3/2{\cal I}$), we have to fix the value of $e$ so that
$e^3 f_\pi$ does not change. This condition yields $e=4.82$ when $f_\pi =
93$ MeV, which implies that the soliton mass becomes so heavy as 1.4 GeV.
We expect that the Casimir energy of fluctuating pions \cite{MK} can
reduce it down to 0.87 GeV. In this work, for a comparison, we fix the
nucleon mass to 940 MeV for all parameter sets. Compared with $f_\pi$=64.5
MeV and $e$=5.45, $ef_\pi$ becomes 1.3 times larger and thus the potential
becomes deeper and narrower by the same factor. This is shown by the dashed
line of Fig. 3. The change in $f_\pi$ alone (Set 2) helps the ground
$\Lambda_c$ mass to come down to 2339 MeV with the $\Lambda^*_c$-$\Lambda_c$
mass difference being 270 MeV.

Another way of improving the results is simply to take a larger $|g_Q|$
value (with putting aside the experimental upper limit on $|g_Q|$ for a
while), which makes the potential deeper. The dotted line in Fig. 3 is
what obtained by varying $g_Q$ and $f_Q/2m_{\Phi^*}$ to $-0.92$ while
keeping $f_\pi$=64.5 MeV (Set 3). Surprisingly, this nearly 20\% change
in coupling constants results in about 50\% enhancement in the binding
energy, while the mass splitting is not so much improved compared with
that of Set 1. By the same way, we take the empirical value for $f_\pi$
and vary $g_Q$ so that the ground $\Lambda_c$ mass becomes close to the
experimental value, which is achieved with $g_Q \sim -0.81$ (Set 4).
This parameter set yields comparable $\Lambda_c$ mass spectrum to the
experiments, which looks quite encouraging. Furthermore, if we break
the heavy quark spin symmetric relation, $f_Q/2m_{D^*}=g_Q$, between
the two coupling constants, we can obtain more realistic mass splitting
between $\Lambda^*_c (\frac12{}^-)$ and $\Lambda^*_c(\frac32{}^-)$. As
an example, we choose $g_Q=-0.70$ and $f_Q/2m_{D^*}=-0.85$ with $f_\pi =
93$ MeV (Set 5). Unfortunately, these coupling constants are not close
to the recent estimates of $-0.2 \sim -0.5$ \cite{CNDB}. We regard this
fact as an indication of the important role of higher order corrections
such as light vector mesons.

In Fig. 4, we present our results on $\Lambda_c$ spectrum (obtained with
parameter Set 5) together with the experimental values and the other model
calculations; SM (Skyrme model with the heavy pseudoscalar mesons only)
\cite{RRS}, QM1 (quark model) \cite{CI}, QM2 \cite{CIK}, and QM3 \cite{KT}.
Our result can compete with the quark model calculations quantitatively.
Especially, one can notice that it becomes much more improved compared with
the first trial, SM \cite{RRS}, in the Skyrme model. One may improve the
result by adjusting all the parameters for the best fit. What we have done
in this work is just to vary two coupling constants around the values given
by the nonrelativistic quark model prediction and the heavy quark symmetric
relation, {\it i.e.}, $f_Q/2m_{\Phi^*} = g_Q = -0.75$. As for the other
parameters, we used the empirical value for $f_\pi$ with $e$ being fixed
by $\Delta$-$N$ mass splitting, and experimental values for the heavy meson
masses.

Given in Fig. 5 are the spectrum of $\Sigma_c$ of our prediction (obtained
with Set 5) and the other model calculations. As a reference line, the
ground state of $\Lambda_c$ is adopted. Our result is again comparable to
the others. However, the splitting between $\Sigma^*_c$ [$\Sigma_c(\frac32^+)$]
and $\Sigma_c$ appears too small compared with the experimental data. Note
that the same is true for the quark models except QM2. It is also
interesting to note that one cannot improve this situation simply by
making the hyperfine constant $c$ larger in any way. By eliminating $c_n$
and ${\cal I}$ in the mass formula (\ref{massform}), we obtain a
model-independent relation
\begin{equation}\label{massrel}
m_{\Sigma^*_c}^{} - m_{\Sigma_c}^{}  =  (m_\Delta^{} - m_N^{})
 - \textstyle\frac32 (m_{\Sigma_c}^{} - m_{\Lambda_c}^{}).
\end{equation}
(The same model-independent mass relation holds in the nonrelativistic
quark model of De R\'{u}jula, Georgi, and Glashow \cite{NRQM}.)
Thus, when our model successfully reproduces all the experimental values for
$m_N^{}$, $m_\Delta^{}$, $m_{\Lambda_c}^{}$, and $m_{\Sigma_c}^{}$,
we get
\begin{equation}
m_{\Sigma^*_c}^{} - m_{\Sigma_c}^{}  \sim 40 \text{ MeV}
\end{equation}
as our best prediction. It is only the half of the value evaluated
with the recent $\Sigma^*_c(2533)$ \cite{SKAT}.

We present the $\Lambda_b$ mass spectrum obtained with this parameter set
in Fig. 6 with the other model calculations. The parameter set used for the
charm baryons does not work well in the bottom sector; Set 5 yields the
ground $\Lambda_b$ mass as 5492 MeV which is $\sim 150$ MeV below the
experimental value. We may repeat the same process of varying the $g_Q$
(with keeping the empirical value 93 MeV for $f_\pi$) to fit the $\Lambda_b$
mass of 5641 MeV. We also expect that the heavy quark symmetry relation
(\ref{hqr}) holds well in the bottom sector. This process leads us to
$g_Q=f_Q/2m_{B^*}=-0.65$ (Set 6). The results with this parameter set
are also given in Fig. 6. The mass splitting ($\sim 180$ MeV) between the
excited $\Lambda^*_b$ and the ground state $\Lambda_b$ appears much smaller
than that of the $\Lambda_c$ given in Fig. 4, while the quark model
calculations
show nearly independent mass splittings whether the heavy constituent
is a $c$-quark ($\sim 370$ MeV) or a $b$-quark ($\sim 330$--$390$ MeV).
Together with the differences in coupling constants fitting the charm
baryons and the bottom baryons, this apparent difference in the mass
splitting is certainly at odds with the {\em heavy quark flavor symmetry.}
Such a heavy quark flavor symmetry is expected to be somehow broken
because of the mass difference between the $c$-quark and $b$-quark.
However, since both are much heavier than the typical scale of the strong
interaction ($\Lambda_{\text{QCD}}\sim 200$ MeV), the actual amount of
the symmetry breaking in nature that occurs at the order of
$\Lambda_{\text{QCD}}/m_Q$ would not be so large.

Such a behavior can be seen also in the $\Sigma_b$ spectrum given in
Fig. 7. Since there is no experimental data for
the $\Sigma_b$ baryons, we can only compare our results with the quark
model predictions. One can find that the mass splitting between the
ground $\Lambda_b$ and the $\Sigma_b (\frac12^+)$ is 180 $\sim$ 190
MeV, which is comparable to the quark model predictions. Also the small mass
splitting ($\sim$ 10 MeV) between $\Sigma_b (\frac12^+)$ and $\Sigma_b
(\frac32^+)$ is still consistent with the quark model predictions. However,
as in the $\Lambda_b$ spectrum, the excitation energy ($\sim$ 170 MeV)
of $\Sigma_b (\frac12^-)$ appears again smaller than the quark model values
($\sim$ 280 MeV).

It may be the ignorance of the soliton-recoil effect in our work that
causes the larger break down of the heavy quark flavor symmetry than
what is actually implied in the model. In order to see this, let us go
back to Fig. 1. We can see that the kinetic effect reduces the binding
energy of the lowest $D$ $(B)$ meson bound state by 410 (240) MeV from
its infinitely heavy mass limit $\Delta\varepsilon_{\infty} = -\frac32
g F^\prime(0) \sim 790$ MeV \cite{OP}. Note that the ratio of the kinetic
effects ($410/240 \sim 1.7$) and the ratio of energy splittings between
the first excited state and the ground state $(\sim 300/200)$ are very
close to the square root of the (inverse) mass ratio ($\sqrt{2.6} \sim
1.6$). One can easily understand this feature in the harmonic oscillator
approximation. Thus, in our working frame, the fact that $B$ mesons are
2.6 times heavier than $D$ mesons becomes directly reflected in the results.
A simple way of estimating the soliton-recoil effect is to use the
``reduced mass" of the soliton--heavy-meson system, as discussed in Refs.
\cite{SS94,OP}. With the soliton mass about 1 GeV, the reduced masses
of the $D$ mesons and $B$ mesons become $\sim 2/3$ GeV and $\sim 5/6$ GeV,
respectively. Then, the use of these small reduced masses can widen the
energy splittings and their small ratio $\sim 5/4$ will not break the
heavy quark flavor symmetry so seriously. (See also Fig. 4 of Ref.
\cite{OP}.) On the other hand, it will require
stronger potentials to overcome the larger kinetic energies, which should
be supplied by including the light vector mesons and/or higher derivative
terms into the Lagrangian \cite{LVM}.

In summary, we have studied the heavy baryon spectrum in the bound state
approach to the Skyrme model by using the exactly-solved heavy meson
bound states of a given Lagrangian. Our results are {\em qualitatively}
and/or {\em quantitatively} comparable to the experimental observations and
the quark model calculations in the charm/bottom sector. The
nearly degenerate doublets in the spectrum are consistent with the heavy
quark {\em spin} symmetry, and our work has a great improvement
compared with the first trial \cite{RRS} of this model. However, the
absence of the soliton-recoil in our framework breaks the heavy quark
{\em flavor} symmetry more than the model really implies. To be consistent
with both the heavy quark spin and flavor symmetry, such a soliton-recoil
effect should be incorporated into the picture.


\acknowledgements

One of us (B.-Y.P) thanks the Institute for Nuclear Theory at the
University of Washington and the Department of Physics and Astronomy
of the University of South Carolina for their hospitality during the
completion of this work.
This work was supported in part by the National Science Council of
ROC under Grant No. NSC84-2811-M002-036 and in part by the Korea
Science and Engineering Foundation through the SRC program.

\appendix

\section*{}

In this Appendix, we present the normalization condition of the heavy
meson fields and the explicit form of the hyperfine constants.
As discussed in Sec. II, the heavy meson fields are normalized
to give a unit heavy flavor number. For a given grand spin $k$ with
parity $\pi = (-1)^{k\pm1/2}$, this condition can be written
explicitly as
\begin{eqnarray}
1 = \int dr r^2 \Bigl\{ &&
2 \varepsilon_n [ |\varphi|^2 + |\varphi_1^*|^2 + \lambda_\pm
|\varphi_2^*|^2 + \lambda_\mp |\varphi_3^*|^2 ] \nonumber \\ &&
\mbox{} + (\varphi_0^{*\prime\dagger} \varphi_1^* + \varphi_1^{*\dagger}
\varphi_0^{*\prime})
+ (\varphi_0^{*\dagger} \varphi_2^* + \varphi_2^{*\dagger}\varphi_0^*)
\upsilon \mu_\pm
+ (\varphi_0^{*\dagger} \varphi_3^* + \varphi_3^{*\dagger}\varphi_0^*)
\left( \frac{1}{r} + \gamma_\mp \upsilon \right) \lambda_\mp
\nonumber \\ && \mbox{}
- g_Q [
   (\varphi_1^{*\dagger} \varphi_2^* + \varphi_2^{*\dagger} \varphi_1^*)
   a_1 \mu_\pm
 + (\varphi_1^{*\dagger} \varphi_3^* + \varphi_3^{*\dagger} \varphi_1^*)
   a_1 \mu_\mp
 + |\varphi_2^*|^2 (a_1+a_2) \mu_\pm
\nonumber \\ && \qquad \mbox{}
 + |\varphi_3^*|^2 (a_1+a_2) \mu_\mp
 + (\varphi_1^{*\dagger} \varphi_3^* + \varphi_3^{*\dagger} \varphi_1^*)
   a_1 \mu_\mp ] \Bigr\},
\end{eqnarray}
where the constants $\lambda_\pm$, $\mu_\pm$, and $\gamma_\pm$ are
written in terms of $k$ as
\begin{equation} \renewcommand\arraystretch{1}
\begin{array}{l}
\lambda_+ = (k-1/2)(k+1/2), \\
\mu_+ = k-1/2, \\
\gamma_+ = \mu_+/\lambda_+ = 1/(k+1/2), \end{array}
\hskip 1.5cm
\begin{array}{l}
\lambda_- = (k+1/2)(k+3/2), \\
\mu_- = -(k+3/2), \\
\gamma_- = \mu_-/\lambda_- = -1/(k+1/2). \end{array}
\end{equation}

The $c$-value defined in Eq. (\ref{theta}) can be written as
\begin{eqnarray} \label{theta1}
\lefteqn{\quad \langle \bar n; k, k_3; \pi=(-1)^{k\pm1/2} \mid \bbox{\Theta}
\mid \bar n; k, k_3; \pi=(-1)^{k\pm1/2} \rangle}  \nonumber \\
&=& c^n_\tau
 {\cal Y}^{(\pm)\dagger}_{kk_3} \bbox{\tau} {\cal Y}^{(\pm)}_{kk_3}
+ c^n_\ell
 {\cal Y}^{(\pm)\dagger}_{kk_3} {\bf L} {\cal Y}^{(\pm)}_{kk_3}
+ c^n_g
 {\cal Y}^{(\pm)\dagger}_{kk_3} {\bf G} {\cal Y}^{(\mp)}_{kk_3}
\\
&\equiv&
-c_{n}
\langle \bar n; k, k_3; \pi=(-1)^{k\pm1/2} \mid {\bf K}_{\text{bf}}
\mid \bar n; k, k_3; \pi=(-1)^{k\pm1/2} \rangle, \nonumber
\end{eqnarray}
where the functionals $c^n_\tau$, $c^n_\ell$, and $c^n_g$
are obtained by inserting Eqs. (\ref{VA}),
(\ref{wf}), and (\ref{tva}) into Eq. (\ref{pth}). Their explicit
expressions are as follows:
\begin{eqnarray}   \label{B1} 
c^n_\tau = \int dr r^2 \Bigl\{ &&
\varepsilon_n (|\varphi|^2 + |\varphi^*_1|^2) (t_1+t_2)
+ \varepsilon_n |\varphi^*_2|^2 [(t_1+t_2)\lambda_\pm + t_2\mu_\pm]
\nonumber \\ && \mbox{}
+ \varepsilon_n |\varphi^*_3|^2 [(t_1+t_2)\lambda_\mp + t_2\mu_\mp]
+ \varepsilon_n (\varphi_2^{*\dagger} \varphi_3^*
              +\varphi_3^{*\dagger} \varphi_2^*) \mu_\pm
\nonumber \\ && \mbox{}
+ {\textstyle\frac12} ( \varphi_0^{*\prime\dagger} \varphi_1^*
          + \varphi_1^{*\dagger} \varphi_0^{*\prime} ) (t_1+t_2)
\nonumber \\ && \mbox{}
+ {\textstyle\frac12} (\varphi_0^{*\dagger} \varphi_2^*
         + \varphi_2^{*\dagger} \varphi_0^*)
 \{ \upsilon [(t_1+t_2) \lambda_\pm + t_2 \mu_\pm ] \gamma_\pm
   + \left( \frac{1}{r} + \upsilon\gamma_\mp \right)\mu_\pm \}
\nonumber \\ && \mbox{}
+ {\textstyle\frac12} (\varphi_0^{*\dagger} \varphi_3^*
         + \varphi_3^{*\dagger} \varphi_0^*)
 \{ \upsilon \gamma_\pm \mu_\pm
  + \left(\frac{1}{r} + \upsilon \gamma_\mp \right)
    [(t_1+t_2)\lambda_\mp + t_2\mu_\mp] \}
\nonumber \\ && \mbox{}
- {\textstyle\frac12} g_Q (\varphi_2^{*\dagger} \varphi_3^*
             + \varphi_3^{*\dagger} \varphi_2^*)
  (a_1+a_2)(t_1+t_2)\lambda_\mp
- {\textstyle\frac12} g_Q (\varphi_2^{*\dagger} k_2 + k_2^\dagger \varphi_2^*)
  t_3 \mu_\mp
\nonumber \\ && \mbox{}
- {\textstyle\frac12} g_Q (\varphi_2^{*\dagger} k_3 + k_3^\dagger \varphi_2^*)
  t_3 \mu_\pm
+ {\textstyle\frac12} g_Q (\varphi_3^{*\dagger} k_2 + k_2^\dagger \varphi_3^*)
  t_3 (2+\mu_\pm)
\nonumber \\ && \mbox{}
- {\textstyle\frac12} g_Q (\varphi_3^{*\dagger} k_3 + k_3^\dagger \varphi_3^*)
  t_3 \mu_\pm \Bigr\},
\end{eqnarray}
\begin{eqnarray}  \label{B2}  
c^n_\ell = \int dr r^2 \Bigl\{ &&
\varepsilon_n |\varphi|^2 t_2 \gamma_\mp
+ \varepsilon_n |\varphi_1^*|^2 (2t_1+t_2) \gamma_\mp
+ \varepsilon_n |\varphi_2^*|^2 t_2 \gamma_\pm (1-\lambda_\pm-\mu_\pm)
\nonumber \\ && \mbox{}
+ \varepsilon_n |\varphi_3^*|^2 [(2t_1+t_2)\mu_\mp-t_2\gamma_\pm\mu_\mp]
+ \varepsilon_n (\varphi_2^{*\dagger} \varphi_3^*
             + \varphi_3^{*\dagger} \varphi_2^*)
  (1-\gamma_\pm\mu_\pm)
\nonumber \\ && \mbox{}
- {\textstyle\frac12} (\varphi_0^{*\prime\dagger} \varphi_1^*
                     + \varphi_1^{*\dagger} \varphi_0^{*\prime})
  (2t_1+t_2) \gamma_\pm
\nonumber \\ && \mbox{}
+ {\textstyle\frac12} (\varphi_0^{*\dagger} \varphi_2^*
                     + \varphi_2^{*\dagger} \varphi_0^*)
  \{ \upsilon t_2 (1-\lambda_\pm-\mu_\pm) \gamma_\pm^2
   + \left( \frac{1}{r} + \upsilon\gamma_\mp \right)(1-\gamma_\pm\mu_\pm) \}
\nonumber \\ && \mbox{}
+ {\textstyle\frac12} (\varphi_0^{*\dagger} \varphi_3^*
                     + \varphi_3^{*\dagger} \varphi_0^*)
  \{ \upsilon \gamma_\pm (1-\gamma_\pm\mu_\pm)
   + \left( \frac{1}{r} + \upsilon\gamma_\mp \right)
     [ (2t_1+t_2) \mu_\mp - t_2 \gamma_\pm \mu_\mp ] \}
\nonumber \\ && \mbox{}
- {\textstyle\frac12} f_Q (\varphi^\dagger \varphi_0^*
                         + \varphi_0^{*\dagger} \varphi)
  t_3 \gamma_\pm
- {\textstyle\frac12} g_Q (\varphi_1^{*\dagger} \varphi_2^*
                         + \varphi_2^{*\dagger} \varphi_1^*)
  a_1 t_1 \gamma_\pm
\nonumber \\ && \mbox{}
- {\textstyle\frac12} g_Q (\varphi_1^{*\dagger} \varphi_3^*
                         + \varphi_3^{*\dagger} \varphi_1^*)
  a_1 t_1 (1+\gamma_\pm)
\nonumber \\ && \mbox{}
- {\textstyle\frac12} g_Q [|\varphi_2^*|^2
                         + |\varphi_3^*|^2 (1+\gamma_\pm)]
  (a_1+a_2)(t_1+t_2)
\nonumber \\ && \mbox{}
- {\textstyle\frac12} g_Q (\varphi_2^{*\dagger} \varphi_3^*
                         + \varphi_3^{*\dagger} \varphi_2^*)
  (a_1+a_2)(t_1+t_2) \mu_\mp
+ {\textstyle\frac12} g_Q (\varphi_1^{*\dagger} k_1
                         + k_1^\dagger \varphi_1^*)
  t_3 \gamma_\pm
\nonumber \\ && \mbox{}
+ {\textstyle\frac12} g_Q (\varphi_2^{*\dagger} k_2
                         + k_2^\dagger \varphi_2^*)
  t_3 (1+\gamma_\pm)(1+\mu_\mp)
\nonumber \\ && \mbox{}
+ {\textstyle\frac12} g_Q (\varphi_2^{*\dagger} k_3
                         + k_3^\dagger \varphi_2^*)
  t_3 (1+\mu_\pm) \gamma_\pm
\nonumber \\ && \mbox{}
- {\textstyle\frac12} g_Q (\varphi_3^{*\dagger} k_2
                         + k_2^\dagger \varphi_3^*)
  t_3 (2+\mu_\pm) \gamma_\pm
\nonumber \\ && \mbox{}
+ {\textstyle\frac12} g_Q (\varphi_3^{*\dagger} k_3
                         + k_3^\dagger \varphi_3^*)
  t_3 [ 1 + \mu_\pm (1+\gamma_\pm) ] \Bigr\},
\end{eqnarray}
\begin{eqnarray}   \label{B3}  
c^n_g = \int dr r^2 \Bigl\{ &&
\varepsilon_n |\varphi|^2 t_2 \gamma_\pm
+ \varepsilon_n |\varphi_1^*|^2 (2t_1+t_2) \gamma_\pm
+ \varepsilon_n |\varphi_2^*|^2
  t_2 [1+\gamma_\mp (1-\lambda_\pm-\mu_\pm)]
\nonumber \\ && \mbox{}
- \varepsilon_n |\varphi_3^*|^2
  [(2t_1+t_2)\mu_\mp+t_2(\gamma_\mp\mu_\mp-1)]
+ \varepsilon_n (\varphi_2^{*\dagger} \varphi_3^*
             + \varphi_3^{*\dagger} \varphi_2^*)
  \gamma_\pm\mu_\pm
\nonumber \\ && \mbox{}
+ {\textstyle\frac12} (\varphi_0^{*\prime\dagger} \varphi_1^*
                     + \varphi_1^{*\dagger} \varphi_0^{*\prime})
  (2t_1+t_2) \gamma_\pm
\nonumber \\ && \mbox{}
+ {\textstyle\frac12} (\varphi_0^{*\dagger} \varphi_2^*
                     + \varphi_2^{*\dagger} \varphi_0^*)
  \{ \upsilon t_2 [1+\gamma_\mp(1-\lambda_\pm-\mu_\pm)] \gamma_\pm
   + \left( \frac{1}{r} + \upsilon\gamma_\mp \right)\gamma_\pm\mu_\pm \}
\nonumber \\ && \mbox{}
+ {\textstyle\frac12} (\varphi_0^{*\dagger} \varphi_3^*
                     + \varphi_3^{*\dagger} \varphi_0^*)
  \{ \upsilon \gamma_\pm^2 \mu_\pm
   - \left( \frac{1}{r} + \upsilon\gamma_\mp \right)
     [ (2t_1+t_2) \mu_\mp + t_2 (\gamma_\mp \mu_\mp - 1) ] \}
\nonumber \\ && \mbox{}
+ {\textstyle\frac12} f_Q (\varphi^\dagger \varphi_0^*
                         + \varphi_0^{*\dagger} \varphi)
  t_3 \gamma_\pm
- {\textstyle\frac12} g_Q (\varphi_1^{*\dagger} \varphi_2^*
                         + \varphi_2^{*\dagger} \varphi_1^*)
  a_1 t_1 (1+\gamma_\mp)
\nonumber \\ && \mbox{}
+ {\textstyle\frac12} g_Q (\varphi_1^{*\dagger} \varphi_3^*
                         + \varphi_3^{*\dagger} \varphi_1^*)
  a_1 t_1 \gamma_\pm
+ {\textstyle\frac12} g_Q |\varphi_3^*|^2
  (a_1+a_2)(t_1+t_2) \gamma_\pm
\nonumber \\ && \mbox{}
- {\textstyle\frac12} g_Q (\varphi_2^{*\dagger} \varphi_3^*
                         + \varphi_3^{*\dagger} \varphi_2^*)
  (a_1+a_2)(t_1+t_2) (1-\mu_\mp)
- {\textstyle\frac12} g_Q (\varphi_1^{*\dagger} k_1
                         + k_1^\dagger \varphi_1^*)
  t_3 \gamma_\pm
\nonumber \\ && \mbox{}
- {\textstyle\frac12} g_Q (\varphi_2^{*\dagger} k_2
                         + k_2^\dagger \varphi_2^*)
  t_3 [\gamma_\pm+\mu_\mp(1+\gamma_\pm)]
\nonumber \\ && \mbox{}
- {\textstyle\frac12} g_Q (\varphi_2^{*\dagger} k_3
                         + k_3^\dagger \varphi_2^*)
  t_3 [\gamma_\pm(1+\mu_\pm)-1]
\nonumber \\ && \mbox{}
+ {\textstyle\frac12} g_Q (\varphi_3^{*\dagger} k_2
                         + k_2^\dagger \varphi_3^*)
  t_3 [1+\gamma_\pm(2+\mu_\pm)]
\nonumber \\ && \mbox{}
- {\textstyle\frac12} g_Q (\varphi_3^{*\dagger} k_3
                         + k_3^\dagger \varphi_3^*)
  t_3 (1+\gamma_\pm) \mu_\pm \Bigr\}.
\end{eqnarray}
The functionals $k_1$, $k_2$, and $k_3$ are defined by
\begin{equation}
{\bf D} \times \bbox{\Phi}^{*\dagger}
= i \{ k_1(r) \, \hat {\bf r} {\cal Y}^{(\pm)}_{kk_3}
     + k_2(r) \, {\bf L} {\cal Y}^{(\mp)}_{kk_3}
     + k_3(r) \, {\bf G} {\cal Y}^{(\pm)}_{kk_3} \},
\end{equation}
which gives
\begin{eqnarray}
k_1 &=& - \varphi_2^* \left( \frac{1}{r} + \upsilon \gamma_\pm \right)
\lambda_\pm - \varphi_3^* \upsilon \mu_\mp,
\nonumber \\
k_2 &=& \varphi_1^* \left( \frac{1}{r} + \upsilon\gamma_\mp \right) -
\left( \varphi_3^{*\prime} + \frac{1}{r} \varphi_3^* \right),
\\
k_3 &=& \varphi_1^* \upsilon \gamma_\pm - \left( \varphi_2^{*\prime} +
\frac{1}{r} \varphi_2^* \right). \nonumber
\end{eqnarray}
To obtain those formulas, we have used the conjugate form of Eq.
(\ref{wf}):
\begin{equation} \renewcommand\arraystretch{1.5}\begin{array}{l}
\displaystyle
\Phi({\bf r},t) = e^{-i\varepsilon t} \varphi^\dagger(r)
{\cal Y}^{(\pm)\dagger}_{kk_3}(\hat {\bf r}), \\
\displaystyle
\Phi^{*}_0({\bf r},t) = - e^{-i\varepsilon t} i \varphi^{*\dagger}_0(r)
{\cal Y}^{(\mp)\dagger}_{kk_3}(\hat {\bf r}), \\
\bbox{\Phi}^{*}({\bf r},t) = e^{-i\varepsilon t}
\left[ \varphi^{*\dagger}_1(r) \,
{\cal Y}^{(\mp)\dagger}_{kk_3}(\hat {\bf r}) \, \hat{{\bf r}}
+ \varphi^{*\dagger}_2(r) \, {\cal Y}^{(\pm)\dagger}_{kk_3}(\hat {\bf r})
\, {\bf L}
- \varphi^{*\dagger}_3(r) \,
 {\cal Y}^{(\mp)\dagger}_{kk_3}(\hat {\bf r}) \, ({\bf G} - 2 \hat{\bf
r}) \right],
\end{array}
\end{equation}
where all the operators act on the {\it right-hand side\/}.

Then from Eq. (\ref{theta1}), we can write $c_n$ as
\begin{eqnarray}   \label{ckpi}
c_{n} &=& \frac{1}{\sqrt{k(k+1)(2k+1)}} \{
c^n_\tau \langle n;k;\pi \| \bbox{\tau} \| n;k;\pi \rangle
+ c^n_\ell \langle n;k;\pi \| {\bf L} \| n;k;\pi \rangle \\ \nonumber
&& \hskip 7em \mbox{}
+ c^n_g \langle n;k;\pi \| {\bf G} \| n;k;\pi \rangle \},
\end{eqnarray}
where the ``reduced matrix elements" are calculated as
\begin{eqnarray}
&&
\langle k; \pi=(-1)^{k+1/2} \| \bbox{\tau} \| k; \pi=(-1)^{k+1/2}
\rangle
= \sqrt{\frac{(k+1)(2k+1)}{k}}, \nonumber \\
&&
\langle k; \pi=(-1)^{k-1/2} \| \bbox{\tau} \| k; \pi=(-1)^{k-1/2}
\rangle
= - \sqrt{\frac{k(2k+1)}{k+1}}, \nonumber \\
&&
\langle k; \pi=(-1)^{k+1/2} \| {\bf L} \| k; \pi=(-1)^{k+1/2}
\rangle
= (k - {\textstyle\frac12}) \sqrt{\frac{(k+1)(2k+1)}{k}}, \\
&&
\langle k; \pi=(-1)^{k-1/2} \| {\bf L} \| k; \pi=(-1)^{k-1/2}
\rangle
= (k + {\textstyle\frac32}) \sqrt{\frac{k(2k+1)}{k+1}}, \nonumber \\
&&
\langle k; \pi=(-1)^{k+1/2} \| {\bf G} \| k; \pi=(-1)^{k+1/2}
\rangle
= - {\textstyle\frac12} (k + {\textstyle\frac32})
    \sqrt{\frac{2k+1}{k(k+1)}}, \nonumber \\
&&
\langle k; \pi=(-1)^{k-1/2} \| {\bf G} \| k; \pi=(-1)^{k-1/2}
\rangle
= {\textstyle\frac12} (k - {\textstyle\frac12})
    \sqrt{\frac{2k+1}{k(k+1)}}, \nonumber
\end{eqnarray}
and the others are zero.

As a specific example, the $c$-values for the $k^\pi = \frac12^\pm$
states are given below:
\begin{eqnarray}
c_{\frac12^+} = \int dr r^2 \Bigl\{ &&
{\textstyle\frac23} \varepsilon_{\frac12^+} [ |\varphi|^2 (-t_1+t_2)
  + |\varphi^*_1|^2 (3t_1+t_2) - 2 |\varphi^*_2|^2 (t_1+t_2) ]
\nonumber \\
&& \mbox{} + {\textstyle\frac13} ( \varphi_0^{*\prime\dagger} \varphi^*_1
                         + \varphi_1^{*\dagger} \varphi_0^{*\prime})
     (3t_1+t_2)
+ {\textstyle\frac23} ( \varphi_0^{*\dagger} \varphi^*_2
                         + \varphi_2^{*\dagger} \varphi_0^{*})
     \upsilon (t_1+t_2)
\nonumber \\ && \mbox{}
+ {\textstyle\frac23} f_Q (\varphi^\dagger \varphi^*_0
                         + \varphi_0^{*\dagger} \varphi) t_3
+ {\textstyle\frac23} g_Q (\varphi_1^{*\dagger} \varphi^*_2
                         + \varphi_2^{*\dagger} \varphi_1^{*})
     [a_1 t_1 + 2 \left( \frac{1}{r} - \upsilon \right) t_3 ]
\nonumber \\ && \mbox{}
- {\textstyle\frac23} g_Q |\varphi^*_2|^2
     (a_1+a_2)(t_1+t_2) \Bigr\},
\end{eqnarray}
and
\begin{eqnarray}
c_{\frac12^-} = \int dr r^2 \Bigl\{ &&
{\textstyle\frac23} \varepsilon_{\frac12^-} [ |\varphi|^2 (3t_1+t_2)
  + |\varphi^*_1|^2 (-t_1+t_2) - 2 |\varphi^*_3|^2 (t_1+t_2) ]
\nonumber \\
&& \mbox{} + {\textstyle\frac13} ( \varphi_0^{*\prime\dagger} \varphi^*_1
                         + \varphi_1^{*\dagger} \varphi_0^{*\prime})
     (-t_1+t_2)
 - {\textstyle\frac23} ( \varphi_0^{*\dagger} \varphi^*_3
                         + \varphi_3^{*\dagger} \varphi_0^{*})
     \left( \frac{1}{r} - \upsilon \right)(t_1+t_2)
\nonumber \\ && \mbox{}
 - {\textstyle\frac23} f_Q (\varphi^\dagger \varphi^*_0
                         + \varphi_0^{*\dagger} \varphi) t_3
 - {\textstyle\frac23} g_Q (\varphi_1^{*\dagger} \varphi^*_3
                         + \varphi_3^{*\dagger} \varphi_1^{*})
     (a_1 t_1 - 2 \upsilon t_3 )
\nonumber \\ && \mbox{}
 - {\textstyle\frac23} g_Q |\varphi^*_3|^2
     (a_1+a_2)(t_1+t_2) \Bigr\}.
\end{eqnarray}



\begin{figure}
\centerline{\epsfxsize=0.60\hsize \epsffile{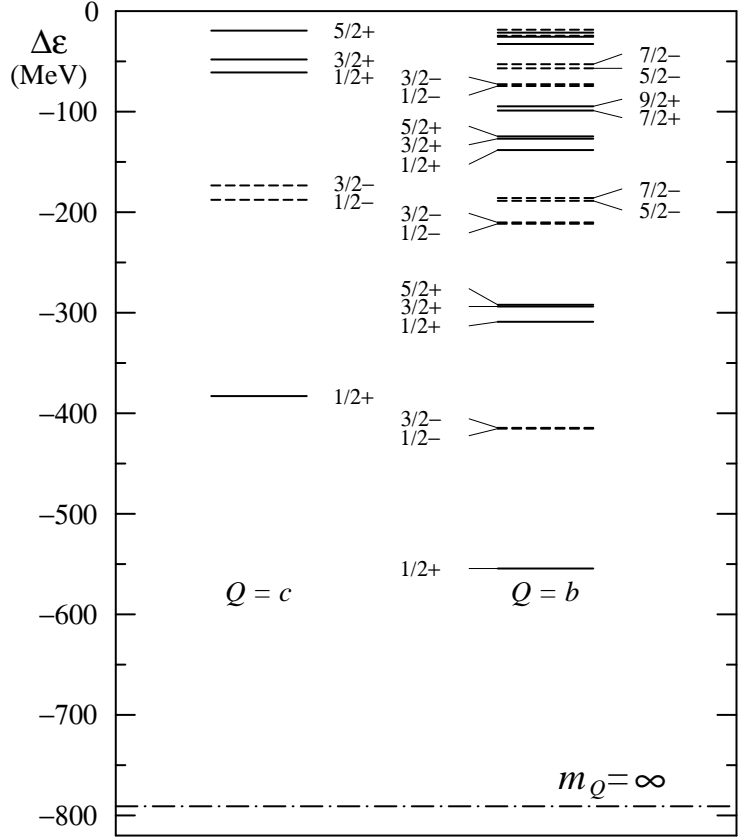}}
\caption{Energy levels of $k^\pi$ bound heavy meson states obtained
with $f_\pi=64.5$ MeV, $e=5.45$, $m_D=1867$ MeV, $m_{D^*}=2010$ MeV,
$m_B^{}=5279$ MeV, $m_{B^*}=5325$ MeV, and
$f_Q/2m_{\Phi^*} = g_Q = -0.75$. The dash-dotted line is the binding energy
obtained in the infinite mass limit.}
\label{f:ene}
\end{figure}

\begin{figure}
\centerline{\epsfxsize=0.60\hsize \epsffile{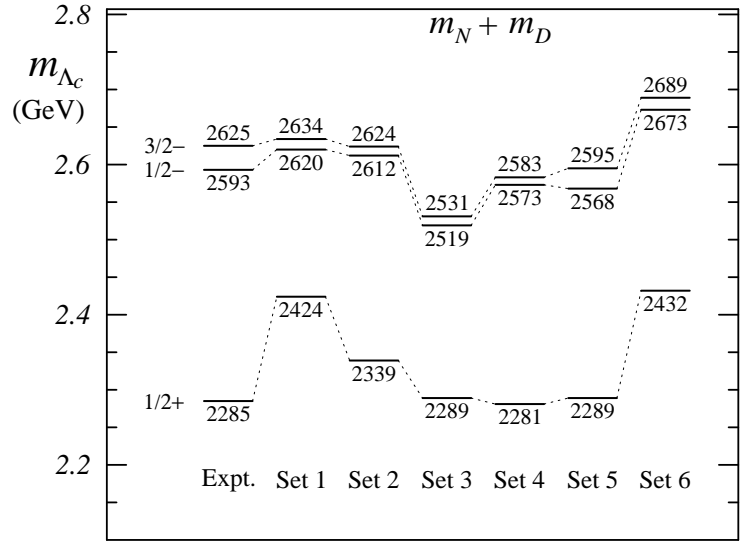}}
\caption{Parameter dependence of $\Lambda_c$ mass spectrum.}
\label{par_lam}
\end{figure}

\begin{figure}
\centerline{\epsfxsize=0.60\hsize \epsffile{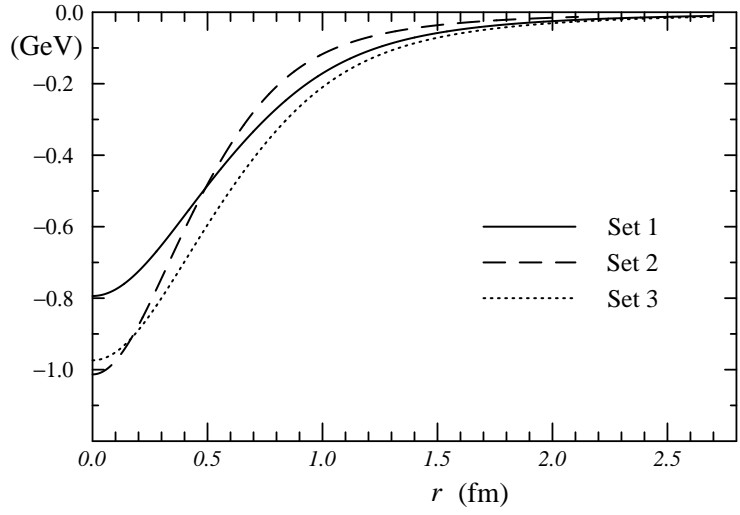}}
\caption{Shape of $\frac12 g_Q^{} [a_1^{}(r)-a_2^{}(r)]$ with
various parameter sets.}
\label{f:pot}
\end{figure}

\begin{figure}
\centerline{\epsfxsize=0.60\hsize \epsffile{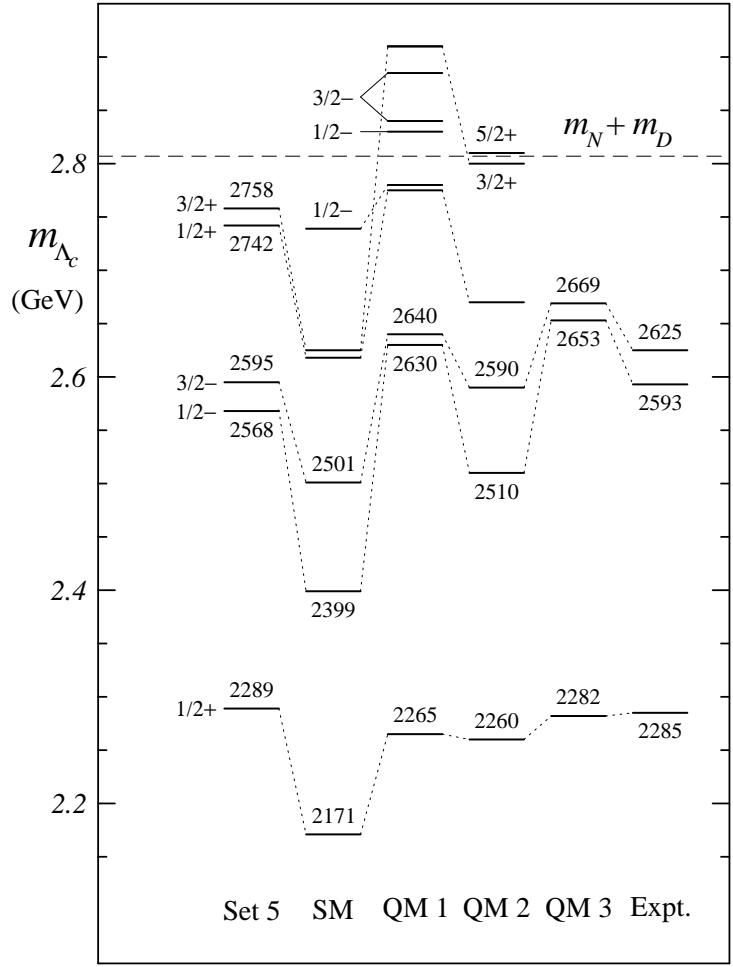}}
\caption{Mass spectrum of $\Lambda_c(j^\pi)$.
The results with Set 5 are presented. For a comparison, we use the
experimental nucleon mass in Set 5. The predictions of other models,
SM (Skyrme Model with only pseudoscalar heavy meson) \protect\cite{RRS},
QM1 (Quark Model) \protect\cite{CI}, QM2 \protect\cite{CIK}, and QM3
\protect\cite{KT} are also given.}
\label{lam_c}
\end{figure}

\begin{figure}
\centerline{\epsfxsize=0.60\hsize \epsffile{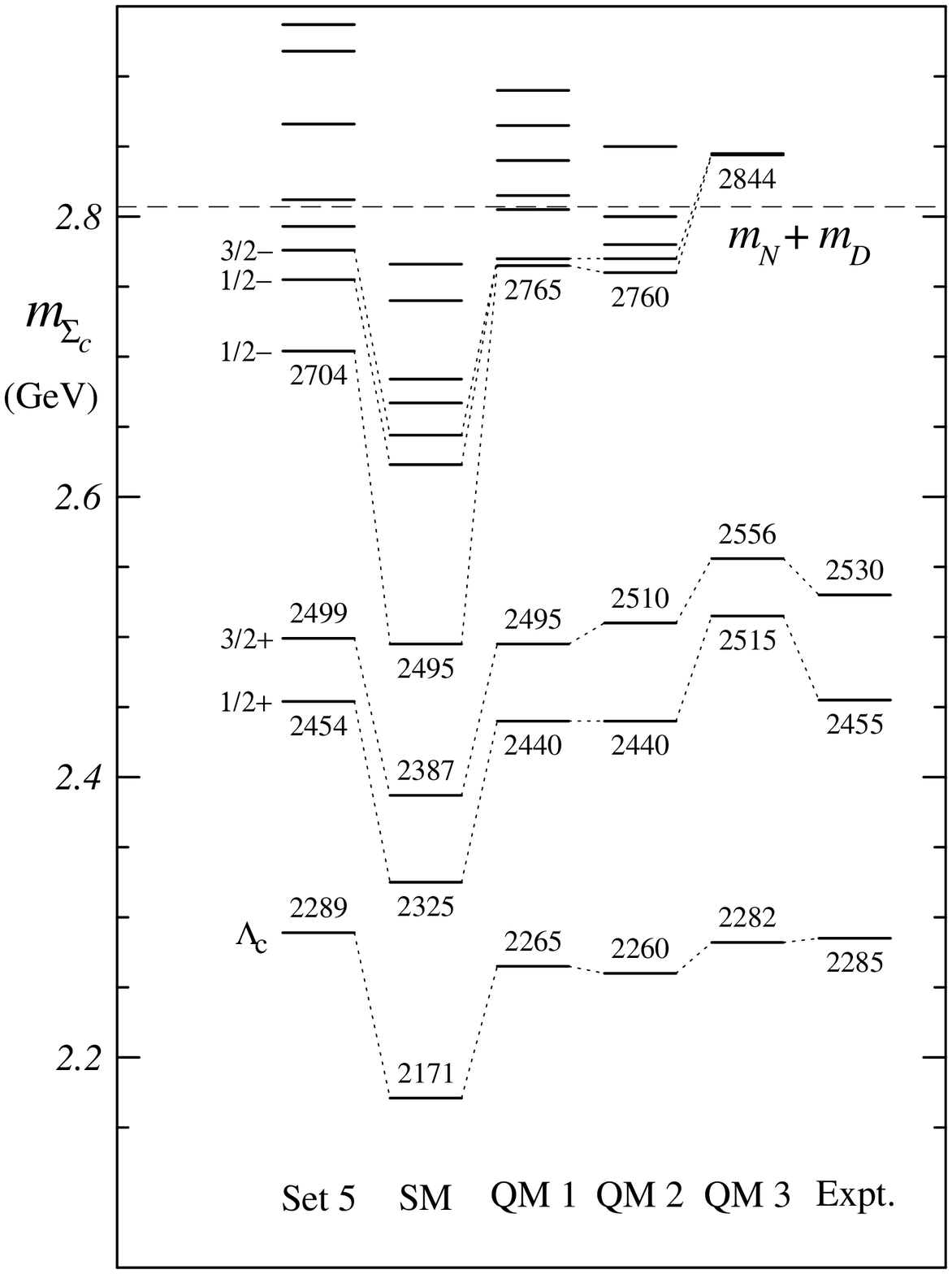}}
\caption{Mass spectrum of $\Sigma_c(j^\pi)$.
Notations are the same as in Fig. \protect\ref{lam_c}.}
\label{sig_c}
\end{figure}

\begin{figure}
\centerline{\epsfxsize=0.60\hsize \epsffile{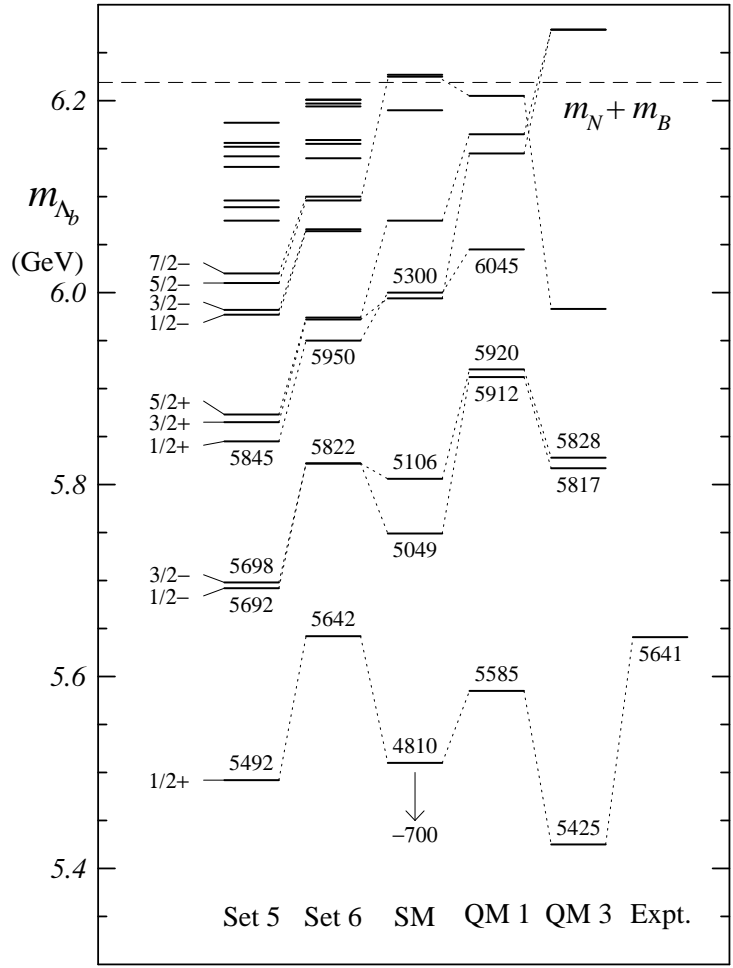}}
\caption{Mass spectrum of $\Lambda_b(j^\pi)$.
The predictions of Set 5 and Set 6 are presented with the results of
SM \protect\cite{RRS}, QM1 \protect\cite{CI}, and QM3 \protect\cite{KT}.}
\label{lam_b}
\end{figure}

\begin{figure}
\centerline{\epsfxsize=0.60\hsize \epsffile{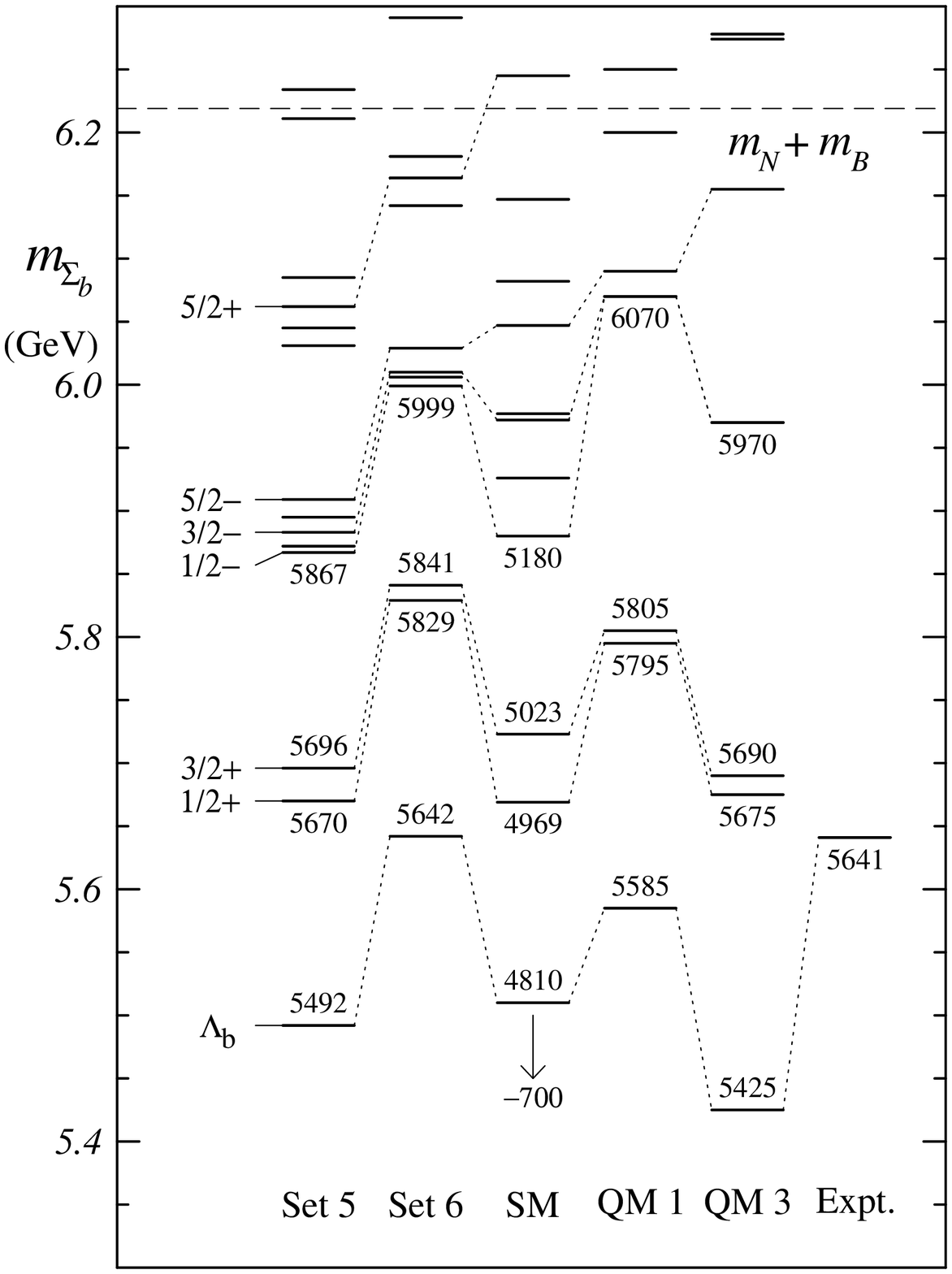}}
\caption{Mass spectrum of $\Sigma_b(j^\pi)$.
Notations are the same as in Fig. \protect\ref{lam_b}.}
\label{sig_b}
\end{figure}


\begin{table}
\caption{Parameter sets. $f_\pi$ is in MeV unit and the others are
dimensionless.}
\begin{tabular}{ccccc}
 & $f_\pi$ & $e$ & $g_Q$ & $f_Q/2m_{\Phi^*}$  \\
\hline
Set 1 & 64.5 & 5.45 & $-0.75$ & $-0.75$ \\
Set 2 & 93.0 & 4.82 & $-0.75$ & $-0.75$ \\
Set 3 & 64.5 & 5.45 & $-0.92$ & $-0.92$ \\
Set 4 & 93.0 & 4.82 & $-0.81$ & $-0.81$ \\
Set 5 & 93.0 & 4.82 & $-0.70$ & $-0.85$ \\
Set 6 & 93.0 & 4.82 & $-0.65$ & $-0.65$ \\
\end{tabular}
\end{table}

\end{document}